\documentclass[twocolumn]{aastex63}

\submitjournal{ApJ}

\shorttitle{Virial radius cold front}
\shortauthors{S. A. Walker et al}
\graphicspath{{./}{figures/}}

\begin{document}

\title{Is there an enormous cold front at the virial radius of the Perseus cluster?}

\correspondingauthor{Stephen Walker}
\email{stephen.walker@uah.edu}

\author{S. A. Walker}
\affiliation{Department of Physics and Astronomy, University of Alabama in Huntsville, Huntsville, AL 35899, USA}

\author{M. S. Mirakhor}
\affiliation{Department of Physics and Astronomy, University of Alabama in Huntsville, Huntsville, AL 35899, USA}

\author{J. ZuHone}
\affiliation{Harvard-Smithsonian Center for Astrophysics, 60 Garden Street, Cambridge, MA 02138, USA}

\author{J. S. Sanders}
\affiliation{Max-Planck-Institute fur extraterrestrische Physik, 85748 Garching, 
Germany}

\author{A. C. Fabian}
\affiliation{Institute of Astronomy, Madingley Road, Cambridge CB3 0HA}

\author{P. Diwanji}
\affiliation{Department of Physics and Astronomy, University of Alabama in Huntsville, Huntsville, AL 35899, USA}




\begin{abstract}
We present new XMM-Newton observations extending the mosaic of the Perseus cluster out to the virial radius to the west. Previous studies with ROSAT have reported a large excess in surface brightness to the west, possibly the result of large scale gas sloshing, but lacked the spatial resolution and depth to determine if this excess lay behind a cold front. In our new XMM observations we have found that there is a sharp edge in X-ray surface brightness near the cluster virial radius (1.7Mpc) to the west, with a width comparable to the mean free path. The temperature measurements obtained with Suzaku data across this edge show that the temperature increases sharply outside this edge. All of these properties are consistent with this edge being the largest cold front ever observed in a galaxy cluster. We also find a second edge to the west, 1.2Mpc from the core, which also appears to be a cold front. Our results indicate that magnetic fields are able to stabilize the cold fronts against turbulence all the way out to the cluster virial radius. By comparing with numerical simulations, we find that these large cold fronts require large impact parameter, low mass ratio mergers which can produce fast gas motions without destroying the cluster core.

\end{abstract}


\section{Introduction}

Cold fronts produced by gas sloshing are commonly seen in the central cores of relaxed cool core galaxy clusters (see \citealt{Markevitch2007}, \citealt{Zuhone2016review} for reviews), where the high surface brightness allows
them to be easily resolved. At these cold fronts, the X-ray surface brightness and gas density drop sharply, while the temperature of the gas rises sharply, the opposite of what occurs for a shock. 

The cold fronts in cluster cores have extremely narrow widths, smaller than the Coulomb mean free path, meaning that processes such as diffusion, conduction and the onset of hydrodynamic instabilities have been heavily suppressed. Magnetic draping, in which the magnetic field is amplified at the cold front edge, is one mechanism believed to be operating to maintain the sharpness of the cold fronts (\citealt{Lyutikov2006}, \citealt{Asai2007}, \citealt{Dursi2007}, \citealt{ZuHone2011}).  

These cold fronts are believed to
be formed due to the sloshing of the cold cluster core as it responds
to the gravitational disturbance created by an infalling subcluster's
dark matter halo during an off-axis minor merger, as
has been simulated by, for example \citet{Tittley2005},
\citet{Ascasibar2006},  \citet{Roediger2011} and \citet{ZuHone2011}. These
simulations predict that the geometric
features of older cold fronts should propagate outwards into
the lower pressure regions of the cluster as they age, producing a characteristic spiral pattern of alternating cold fronts on opposite sides of the cluster at ever greater distance from the cluster core. Exactly how far out this process can operate is uncertain, as the X-ray surface brightness decreases rapidly with radius in galaxy clusters, making detailed studies of sloshing into the cluster outskirts challenging (\citealt{Walker2019}).

\begin{figure*}
\begin{center}

\vbox{
\includegraphics[width=1.0\textwidth]{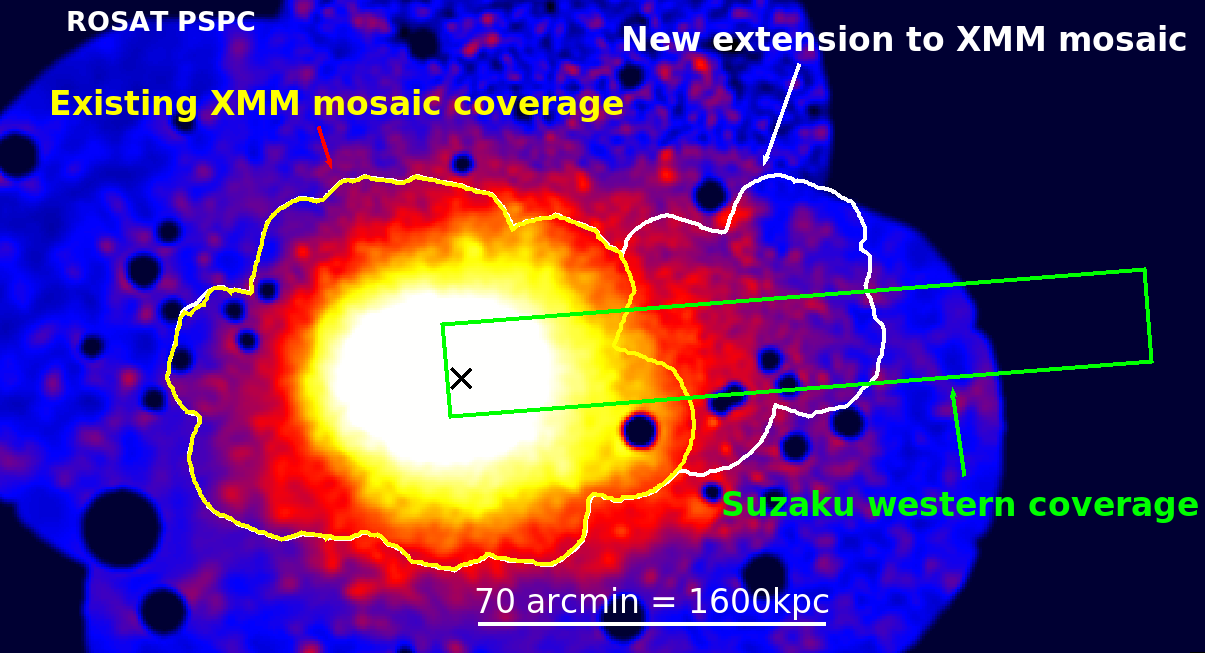}
}

\end{center}
\caption{Large area, shallow ROSAT PSPC mosaic of the Perseus cluster. Overlaid on this is the field that XMM-Newton has covered previously (yellow contour), which we extend with our new observations towards the west (the white contour). The Suzaku coverage to the west is shown by the green area. The location of the core is donated by the black cross.}
\label{ROSATimage}

\end{figure*}

One remarkable recent development in the study of cold fronts is the discovery
of large scale cold fronts reaching out to very large radii, far outside the
cooling radius. In the Perseus cluster, \citet{Simionescu2012} and \citet{Walker2018Nat} have found a cold
front 700kpc from the core to the east (nearly half the virial radius), using XMM and Chandra data. Perseus displays the characteristic sloshing spiral pattern of alternating cold fronts on either side of the cluster, extending from the sloshing activity in the core \citep{Walker2017,Sanders2020} into the cluster outskirts. Similar large scale cold fronts reaching out 
to $\sim$0.5$r_{200}$ have since
been found in just a handful of other clusters, all at much higher redshifts: Abell 2142
(\citealt{Rossetti2013}), RXJ2014.8-2430 (\citealt{Walker2014}) and Abell 1763 (\citealt{Douglass2018}). In all of these
systems, there is a large scale spiral pattern of concentric cold fronts on
opposite sides of the cluster at increasing radii,
suggesting that the structure is one continuous outwardly moving sloshing
motion.

These large scale cold fronts are much older than the cold fronts commonly found
in cluster cores, as they have risen outwards and grown with time. They are over
an order of magnitude further out from the core, and so if we assume a constant
velocity, would be expected to be an order of magnitude older. Based on numerical simulations of rising cold fronts (\citealt{ZuHone2011}), the age of the
a cold front to the East at 700kpc from Perseus's core explored in \citet{Walker2018Nat} should be around 5Gyr, while the age of the cold front which has reached the virial radius is expected to be at least 9 Gyr. Because of this, diffusion processes will have had much longer to
broaden the cold front edge, while instabilities such as Kelvin-Helmholtz
instabilities (which we have found in observations closer to the core, \citealt{Walker2017}) will have had longer to grow.  We would therefore expect the
broadening and substructure behind large scale cold fronts to be significantly
more evolved than that behind young cold fronts in cluster cores, and
consequently easier to resolve spatially.

As cold fronts rise from the core to the outskirts they pass through
different strata of the ICM, which are expected to have
different dominating physics. In the outskirts the physics of the ICM begins to
be dominated by the accretion of gas onto the cluster from large scale structure
filaments and infalling subgroups. Turbulence and bulk motions from the ongoing
formation of the cluster is expected from simulations (\citealt{Lau2009}) to
increase in the outskirts. The rising cold front therefore experiences
vastly different areas of the ICM as it grows, which will have an
influence on the broadening of the cold front and the development of structure
behind it. 

\citet{Simionescu2012} also found evidence from the mosaiced ROSAT PSPC data of Perseus that the sloshing spiral continued even further out to the west, extending out to nearly the virial radius ($r_{200}$=1800kpc=80 arcmin for Perseus, \citealt{Urban2014}). However the low effective area, and large off-axis PSF of ROSAT made it impossible to determine whether this large excess in X-ray surface brightness to the west lay behind a cold front.

To remedy this, in AO17 we extended the XMM mosaic to reach out to the virial radius in a narrow strip to the west of Perseus to see if there is a cold front there.

Many outstanding questions remain about the way cold fronts evolve out to large
radius. How far out can magnetic draping continue to suppress transport
processes, and are we beginning to see this process break down? At some point in the cluster outskirts the cluster magnetic field must interact with that of the surrounding cosmic web filaments \citep{Walker2019}. These new XMM-Newton observations allow us to probe cluster astrophysics into extreme, previously unexplored regimes, providing crucial insights into the interface between clusters and the surrounding cosmic web. 

In this paper we present these new XMM-observations. 
We use a standard $\Lambda$CDM cosmology with $H_{0}=70$  km s$^{-1}$
Mpc$^{-1}$, $\Omega_{M}=0.3$, $\Omega_{\Lambda}$=0.7. All errors unless
otherwise stated are at the 1 $\sigma$ level.

\begin{figure*}
\begin{center}

\vbox{
\includegraphics[width=0.7\textwidth]{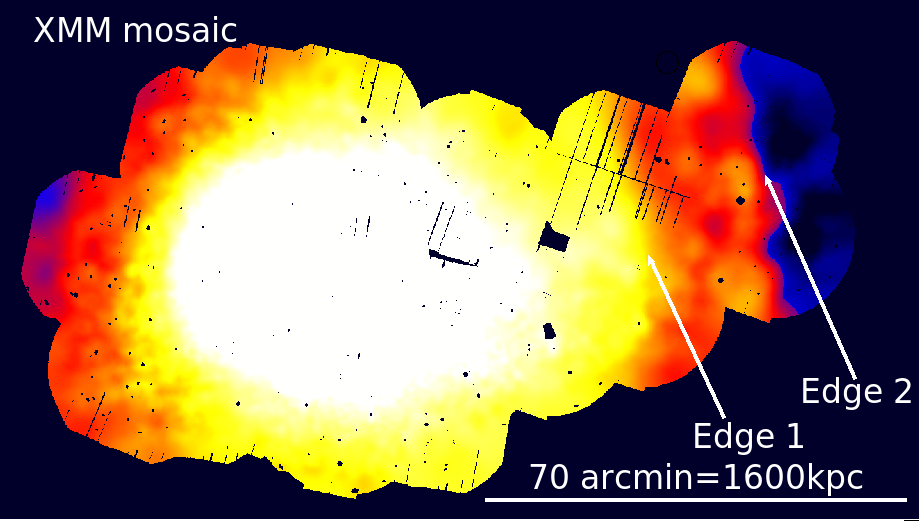}
\includegraphics[width=0.7\textwidth]{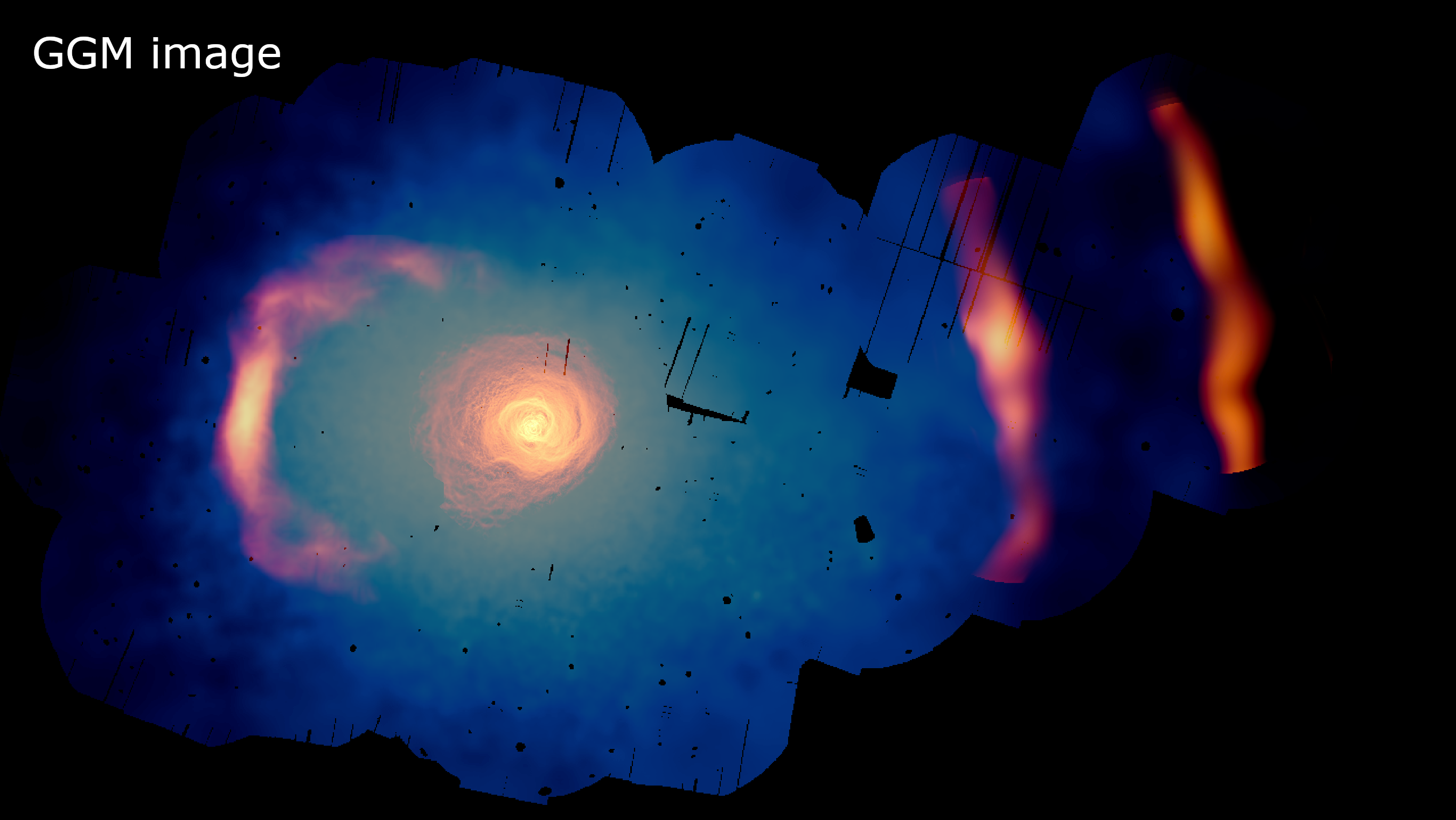}
}

\end{center}
\caption{\emph{Top}: shows the soft band (0.7-1.2keV), background subtracted and exposure corrected XMM mosaic. The two edges to the west are highlighted with arrows. 
\emph{Bottom}: Overplotting the Gaussian Gradient magnitude gradient map (orange) over the normal XMM mosaic (blue) to highlight how these new edges relate to the known cold front structure in the core.}
\label{XMMimage}

\end{figure*}

\section{Data}

Fig. \ref{ROSATimage} shows the wide scale ROSAT PSPC mosaic image of Perseus. Over this we plot the existing regions covered by XMM (the yellow contour) and the region covered by our new extension of the mosaic reaching out to the virial radius to the west (white contour). This new XMM coverage to the west extends over the surface brightness excess found in \citet{Simionescu2012}. We also show the Suzaku coverage in this western direction (green contour). The other 7 strips of the Suzaku mosaic are not shown as none of them overlap with the new XMM data. 

\subsection{XMM-Newton image analysis}

The XMM-Newton observations used are tabulated in table \ref{obsdata}, ordered with the latest observations at the top. Our new observations to extend the mosaic out to the virial radius to the west are 0820720101, 08207202101, 08207201301 and 0820720401.

All of the XMM observations were reduced using the XMM Extended Source Analysis Software \citep{Snowden2008}. The images and exposure maps were extracted in the 0.7-1.2keV band (which has been found to maximize the signal to noise in the outskirts by e.g. \citealt{Tchernin2016}) using \textsc{mos-spectra} and \textsc{pn-spectra}, while particle background images were produced using \textsc{mos-back} and \textsc{pn-back}. Residual soft proton contamination was modelled using spectral fitting and images were produced using the task \textsc{proton}. Point sources were identified using \textsc{cheese} (and also using the Chandra tool \textsc{wavdetect}) and removed. The resulting background subtracted, exposure corrected mosaic image in the 0.7-1.2keV band is shown in the top panel of Fig. \ref{XMMimage}. Two edges are clearly visible, one at 1.7Mpc from the core and the second 1.2Mpc from the core, and these are labelled in the top panel of Fig. \ref{XMMimage}.

To emphasize the edges, and show how they relate to the central cold fronts, we use Gaussian Gradient Magnitude filtering (\citealt{Sanders2016b}, \citealt{Walker2016}), which convolves the image with a Gaussian kernel and finds the gradient on the spatial scale of the kernel. The GGM filtered image is shown in the bottom panel of Fig. \ref{XMMimage} (orange regions), where is is overlaid on the unfiltered XMM mosaic image (blue). For the central 200kpc region we use the Chandra observations in the mosaic.

A local background field (observation 0672770101) lying beyond the virial radius (92 arcmin from the core) to the East is available for a local background measurement, which is subtracted when measuring the X-ray surface brightness. When extracting surface brightness profiles in the outskirts of clusters, it is important to consider possible biases introduced by gas clumping, which become increasingly important near the virial radius \citep{Walker2019}. \citet{Eckert2015} have found that these clumping biases can be overcome by Voronoi tesselating the data and then computing the median surface brightness value in the tesselation in each radial bin. 

\begin{figure*}
\begin{center}

\hbox{\includegraphics[width=0.5\linewidth]{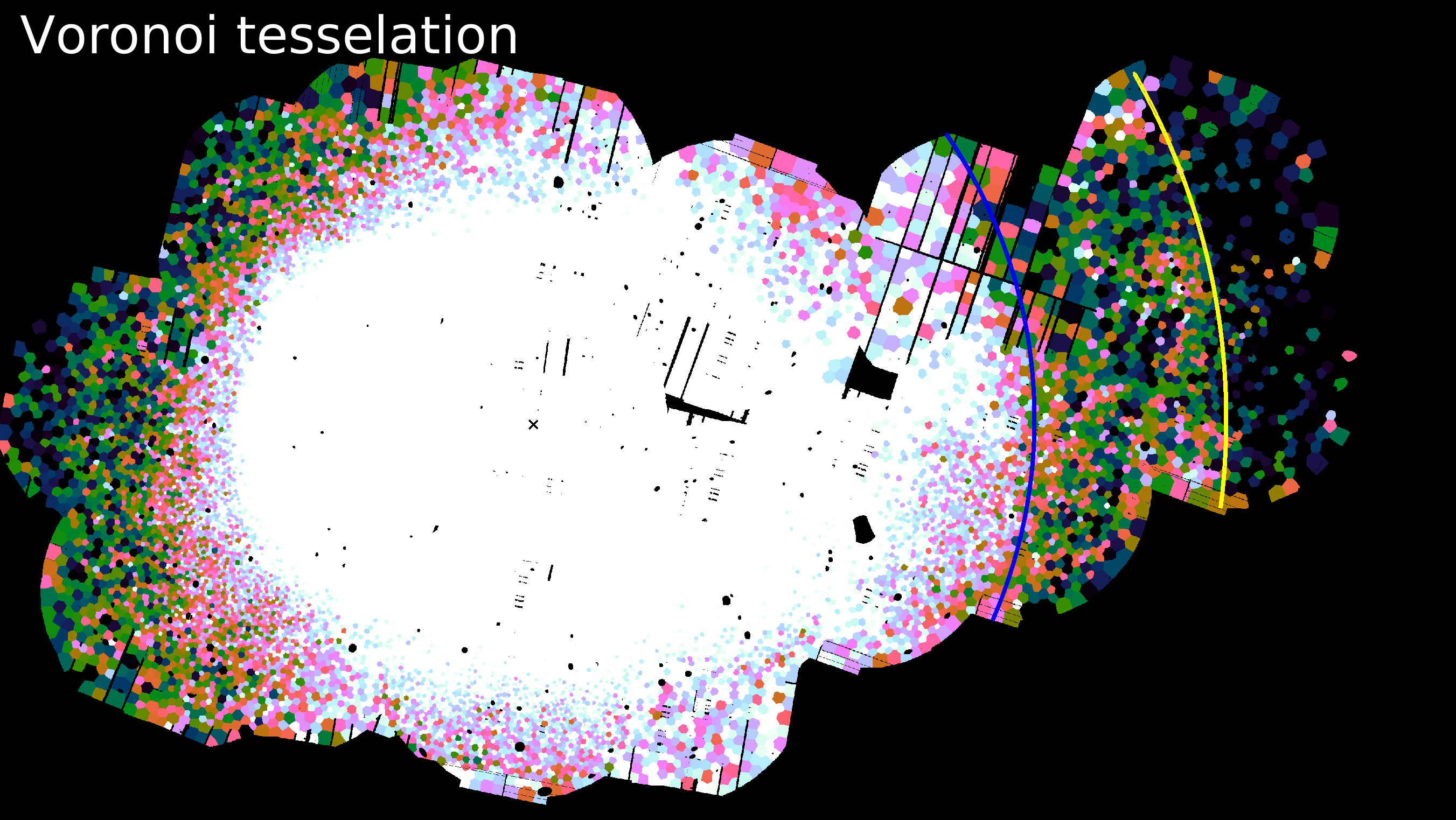}
\hspace{1.0cm}
\includegraphics[width=0.43\linewidth]{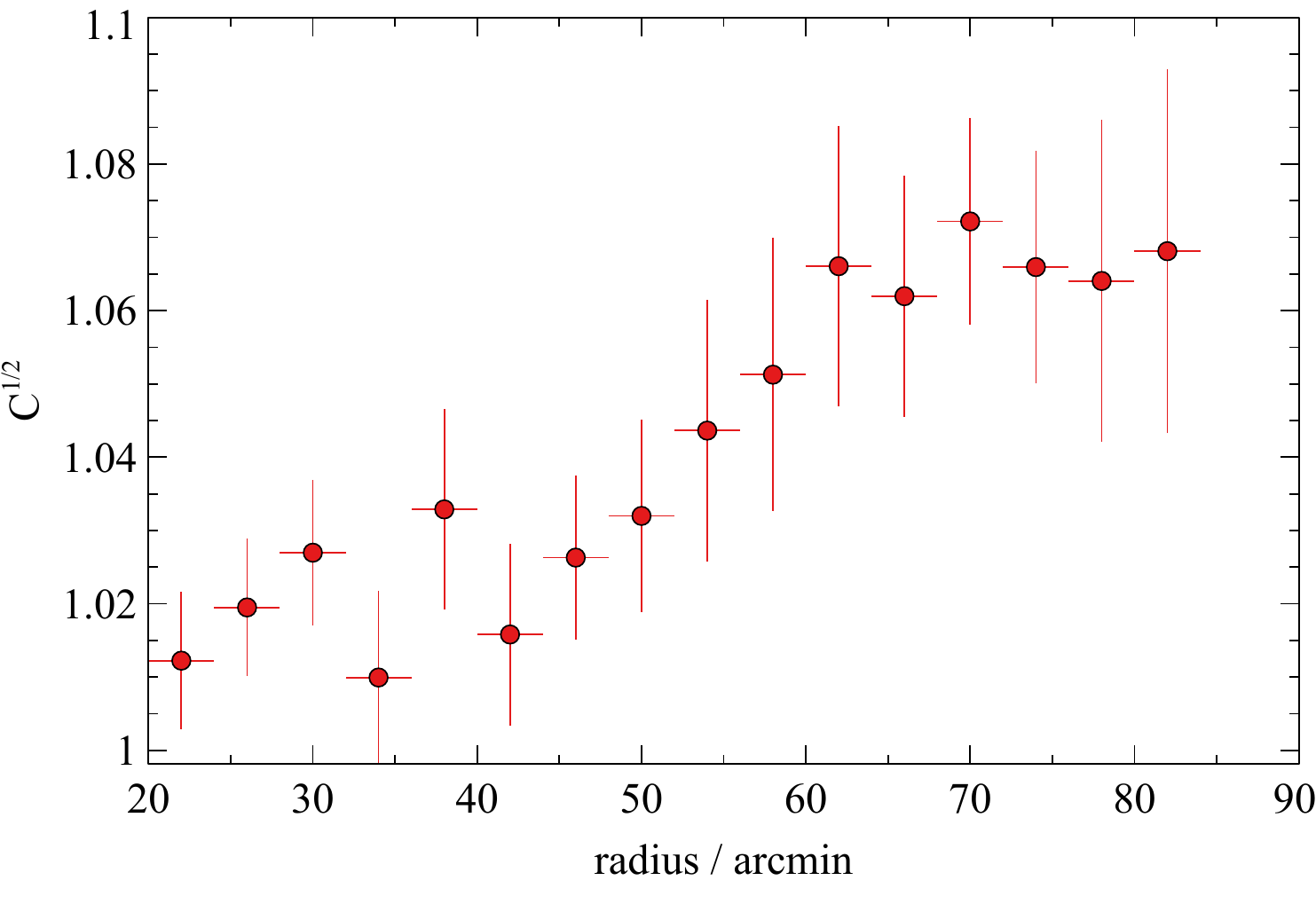}
}

\end{center}
\caption{Left: Voronoi tesselation of the background subtracted, exposure corrected XMM-Newton mosaic. The locations of the two new edges are marked by the blue and yellow curves. Right: Profile of the gas clumping factor infered as the ratio between the mean and the median deprojected density in each annulus of the Voronoi tesselation.
  }
\label{Voronoi_image}

\end{figure*}

\begin{figure*}
\begin{center}

\includegraphics[width=1.0\textwidth]{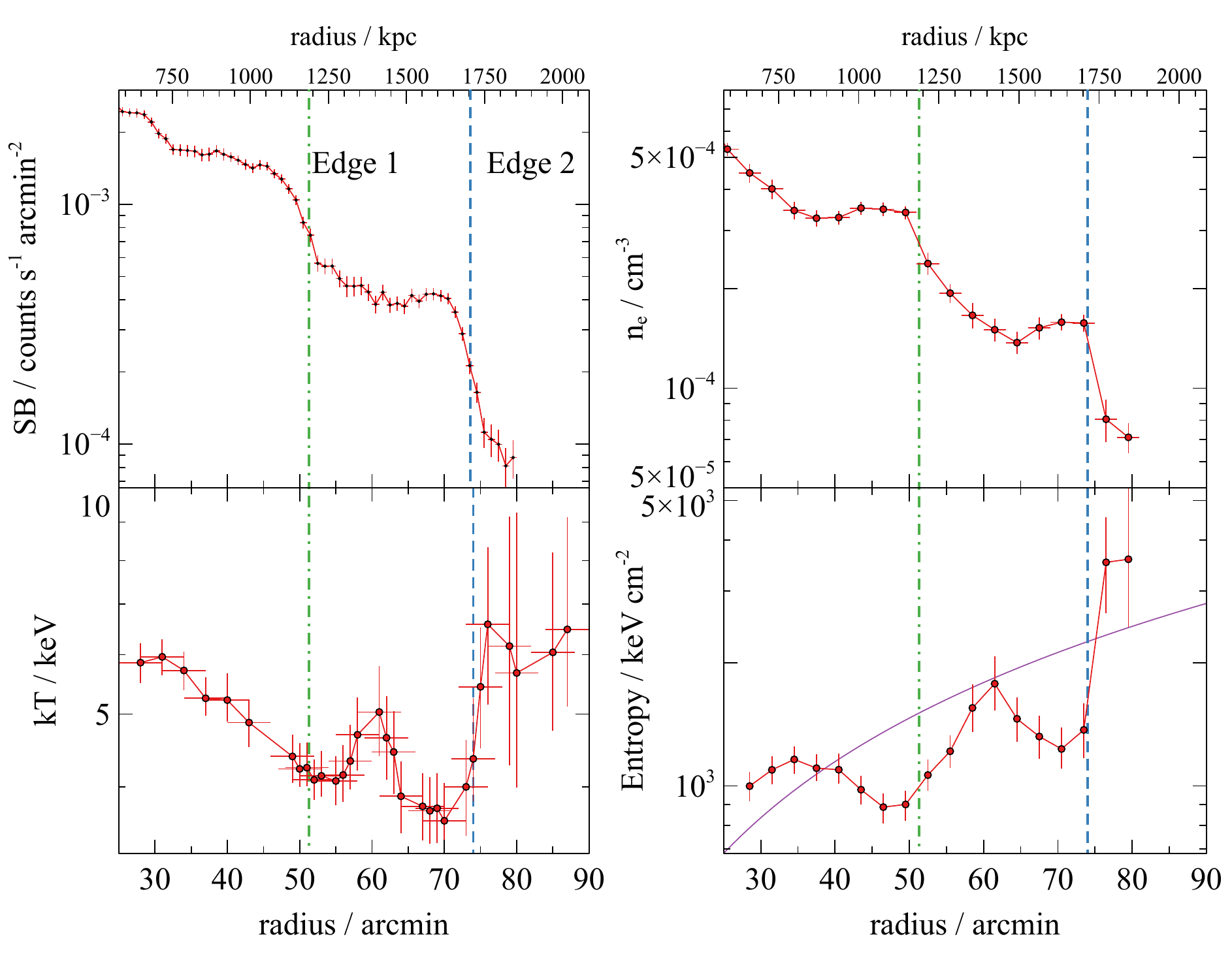}

\end{center}
\caption{Top left: projected surface brightness profile. This is deprojected to produce the deprojected density profile (top right). Bottom left: Temperature profile obtained from the Suzaku data in the same western direction as the XMM observations. Bottom right: Entropy profile compared to the baseline entropy profile from \citealt{Voit2005} (solid purple curve). The dot-dashed and dashed vertical lines show the locations of the two edge determined by fitting the surface brightness profiles with a broken powerlaw model (Fig. \ref{edge_fitting}).
  }
\label{thermo_plots}

\vspace{-0.3cm}
\end{figure*}

We therefore produce a Voronoi tesselation of the XMM mosaic using the method of \citet{Diehl2006}, which is shown in Fig. \ref{Voronoi_image}, where the image is binned so that each region contains at least 20 photons. We then extract the median surface brightness profile in a sector to the west, which is shown in the top left hand panel of Fig. \ref{thermo_plots}. The inferred clumping factor $\sqrt{C}$, which is the square root of the ratio of the mean deprojected surface brightness to the median deprojected surface brightness in each annulus region of the Voronoi tesselation, is plotted in the right hand panel of Fig. \ref{Voronoi_image}. The clumping values are relatively low, reaching up to 1.08, and consistent with those found in other works using the same technique (\citealt{Eckert2015}, \citealt{Tchernin2016} and \citealt{Ghirardini2018}). This clumping bias is therefore very mild and acts just to raise the observed densities, with no significant effect on the shape of the surface brightness profile and the magnitude of the edges we identify. 

The two clear edges visible both in the images and in the median surface brightness profile are marked on Fig. \ref{thermo_plots}. The outer most edge lies at 1.7Mpc from the core, just slightly smaller than the virial radius. The second edge is 1.2Mpc from the core, and runs roughly parallel to the outer edge.

For temperatures above 2keV, the X-ray surface brightness can be converted in the gas density in a way that is largely independent of the metal abundance of the gas (e.g. \citealt{Ghirardini2018}). The deprojected median density profile is shown in the top right hand panel of Fig. \ref{XMMimage}, showing the two sharp drops in gas density at the edges.   

\begin{figure*}
\begin{center}

\hbox{
\includegraphics[width=0.5\textwidth]{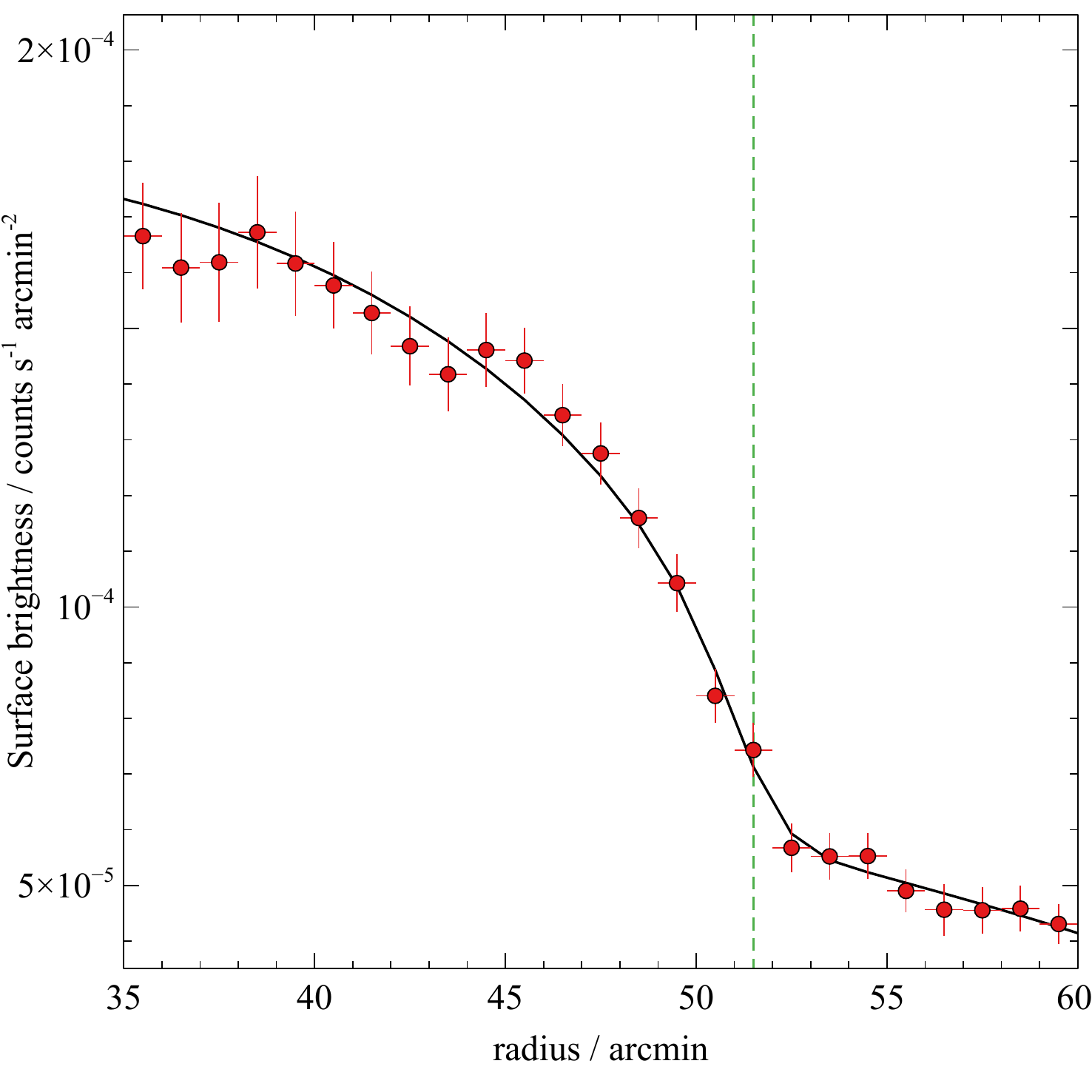}
\includegraphics[width=0.5\textwidth]{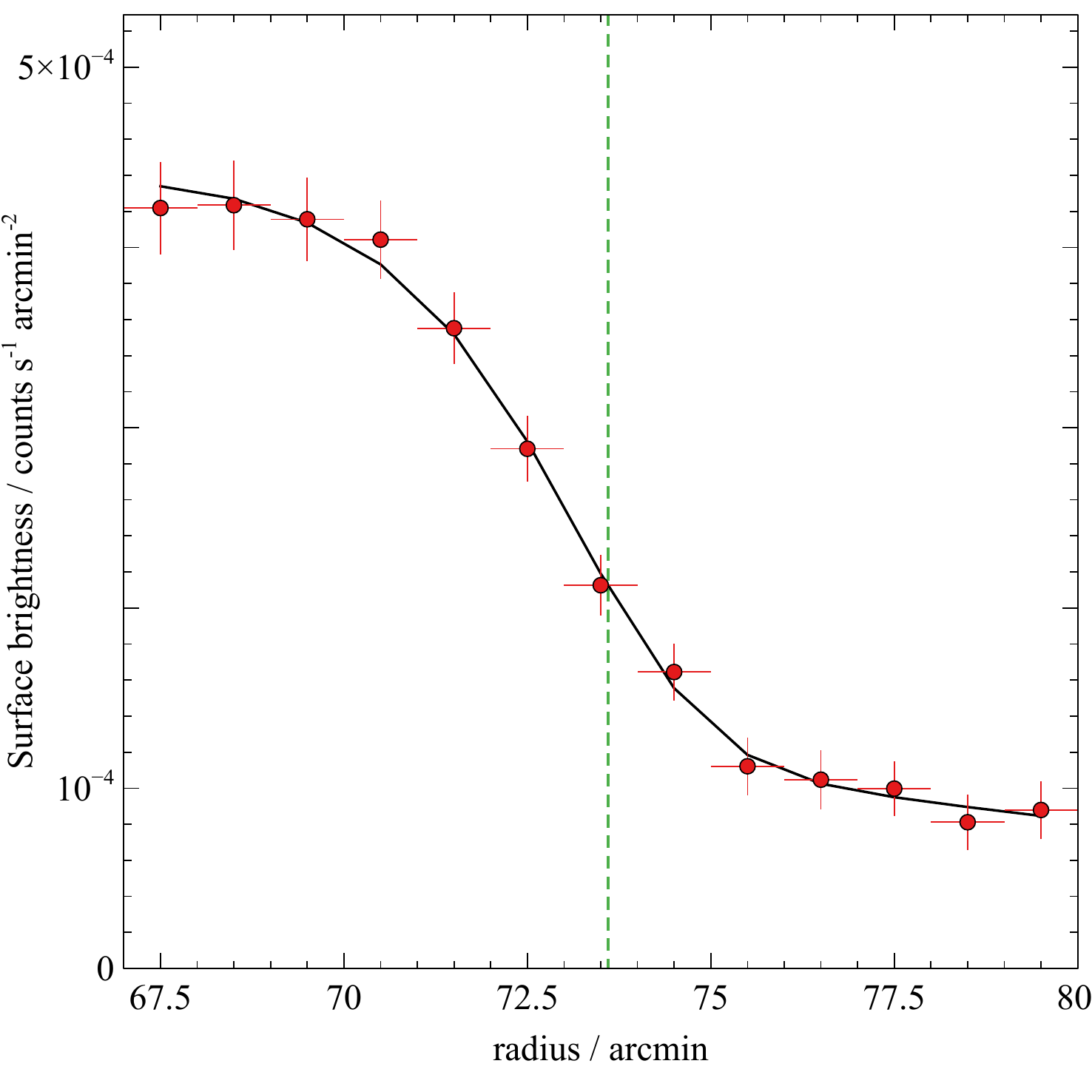}

}

\end{center}
\caption{Fitting the surface brightness profiles of the 50 arcmin radius (left) and 70 arcmin radius (right) edges with a by a model in which the surface brightness jumps from one powerlaw to another, integrated along the line of sight and convolved with the XMM-Newton PSF.
  }
\label{edge_fitting}

\end{figure*}

\subsection{Suzaku spectral analysis}
The high and variable background of XMM-Newton makes it difficult to obtain reliable temperature measurements outside of $r_{500}$ using spectral fitting. We therefore use the Suzaku data to the west (the coverage is shown by the green box in Fig. \ref{ROSATimage}) to obtain temperature measurements. Suzaku's low Earth orbit provided it with the low and stable background needed for accurate X-ray spectroscopy out to the virial radius. These data have previously been explored in \citet{Urban2014} and \citet{Simionescu2012}, however the large PSF of Suzaku prevented and X-ray surface brightness edges from being measured. The Suzaku datasets in the strip to the west are tabulated in table \ref{obsdata_Suzaku}. 

Our XMM-Newton data allow point sources to be identified to a much lower threshold flux than was possible in \citet{Urban2014} (which just used Suzaku data), reaching down to $10^{-14}$ erg cm$^{-2}$ s$^{-1}$. As described in \citet{Walker2013_Centaurus} this reduces the uncertainty in the modelling of the unresolved cosmic X-ray background. In the Suzaku spectral fitting, we follow the methods described in \citet{Walker2013_Centaurus}, in which the contributions from the resolved point sources are included in the background model. All of the parameters in the background model were found using the outermost Suzaku data as in \citet{Urban2014}. The soft X-ray background was modelled with three thermal components corresponding to the Galactic halo, the local hot bubble, and a potential 0.6keV component. The cosmic X-ray background was modelled as a powerlaw with index 1.4 whose normalization was calculated by integrating the cumulative distribution of point sources from \citet{Moretti2003} down to the threshold flux to which we resolve point sources.  To include the systematic uncertainties in the background modelling, we followed \citet{Walker2013_Centaurus} and performed 1000 iterations of the spectral fits, varying all of the background model parameters simultaneously through their possible ranges. The systematic uncertainty in the background model is therefore folded through and it is included in the error bars shown for the temperature profile.    

We extract spectra in a 6 arcmin wide annular region (wide enough to contain sufficient photons, 1000, for a good temperature measurement), which we move along the length of the strip of Suzaku observations 1 arcmin at a time in the regions where the edges are seen by XMM. The resulting temperature profile is shown in the bottom left hand panel of Fig. \ref{thermo_plots}. We see that at each edge the temperature increases sharply as the density drops sharply, exactly what would be expected for cold fronts. 

We combine the Suzaku temperature measurements with the deprojected XMM density measurements to obtain the entropy profile $K=kT/n_{e}^{2/3}$, shown in the bottom right hand panel of Fig. \ref{thermo_plots}. When compared to the baseline entropy profile of \citet{Voit2005} (solid purple curve), the two cold fronts are evident as depressions in the entropy due to the higher density and lower temperature of the gas behind them.

\section{Analysis of the edges}

 We fit the surface brightness profiles of the two edges with a model consisting of a powerlaw jumping to another powerlaw, integrated along the line of sight and convolved by XMM's PSF. This is the same model used in \citet{Sanders2016}, and allows us to fit for the radius of the edge, its width (which is deconvolved from the XMM PSF) and the jump height. The best fits are shown in Fig. \ref{edge_fitting}.
 
 For the inner edge, the best fit radius for the edge is 51.3 arcmin (1.2Mpc), the width is 52 arcseconds (20 kpc), and the density decreases by a factor of 1.4 from 3.4$\pm0.15\times10^{-4}$ cm$^{-3}$ to 2.4$\pm0.15\times10^{-4}$ cm$^{-3}$.
 
 For the outer edge, the best fit radius for the edge is 73.6 arcmin (1.7Mpc), the width is 88 arcsec (34 kpc) and the density decreases by around a factor of 2 from 1.6$\pm0.1\times10^{-4}$ cm$^{-3}$ to 0.8$\pm0.15\times10^{-4}$ cm$^{-3}$.

The density decrease for the outer 1.7Mpc edge is therefore significantly larger than that of the 1.2Mpc edge. The temperature jump is also considerably larger: for the 1.7Mpc it jumps by a factor of $1.7^{+0.3}_{-0.2}$ from 3.8$^{+0.5}_{-0.5}$keV just inside the edge to 6.6$^{+1.7}_{-1.4}$keV just outside it, while for the 1.2Mpc the temperature jump is milder, jumping by a factor of 1.25$^{+0.2}_{-0.15}$ from 4.0$^{+0.3}_{-0.3}$keV to 5.0$^{+0.7}_{-0.5}$keV. The pressure over each edge is therefore continuous, as is typically found for cold fronts.

\begin{figure}
\begin{center}

\includegraphics[width=\linewidth]{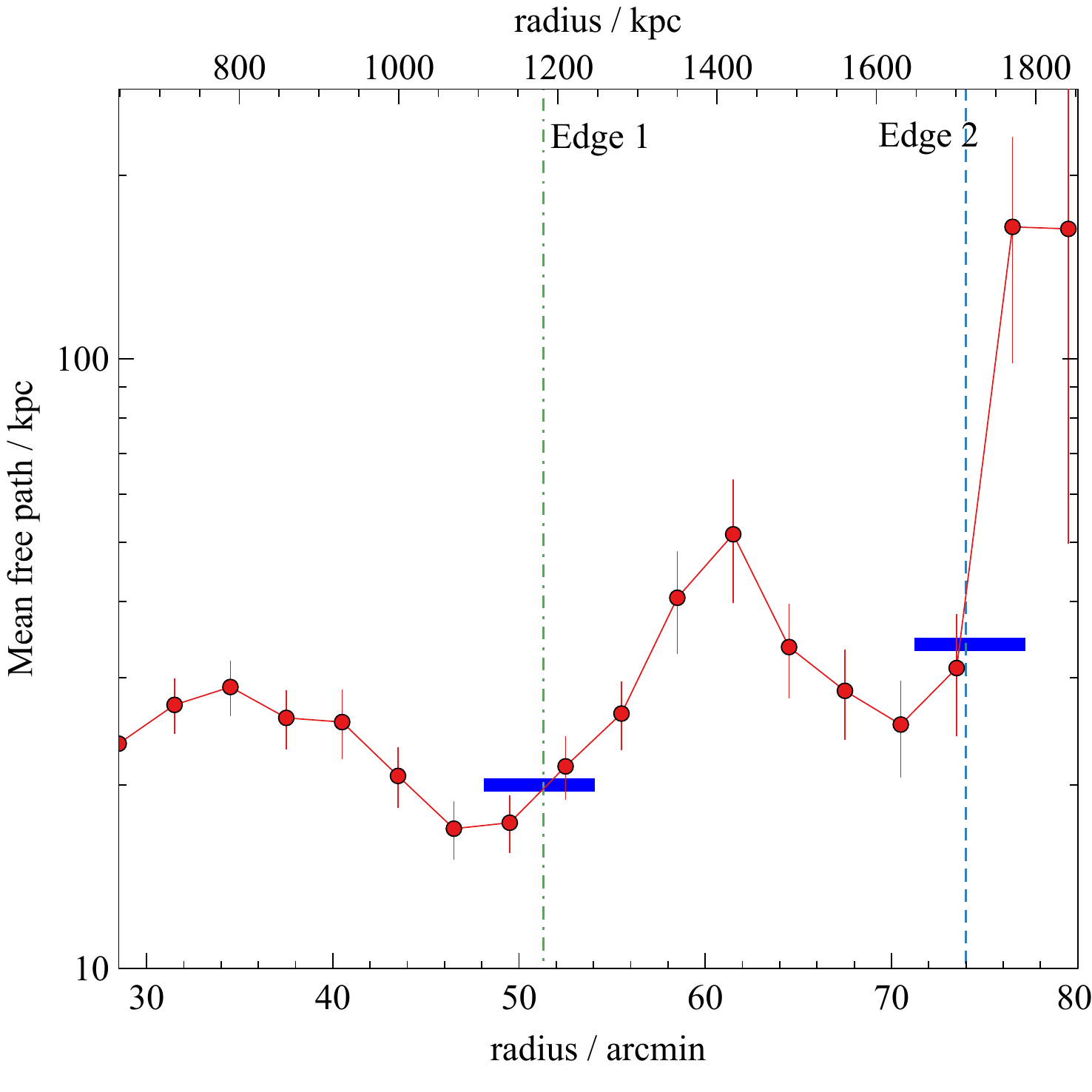}

\end{center}
\caption{Radial profile of the Coulomb mean free path (red points) to the west. The widths of the edges measured by the surface brightness profile fitting are shown as the horizontal blue lines. In both cases the edge width is comparable to the Coulomb mean free path.
  }
\label{cmfp}

\end{figure}

We use the deprojected median density profile and the Suzaku temperature profile to plot the profile of the Coulomb mean free path in Fig. \ref{cmfp}. The widths of the two edges are plotted on Fig. \ref{cmfp} as the horizontal blue lines. We see that the widths of both edges are consistent with the Coulomb mean free path at their radii.

\section{Comparison to simulations}

Many previous simulation studies of sloshing cold fronts have focused on the core region, given that the low-entropy dense gas in the cluster core produces the first and most obvious cold fronts which are formed \citep[e.g.][]{Ascasibar2006,Roediger2011,Roediger2012,Roediger2013,ZuHone2010,ZuHone2011,ZuHone2013b,ZuHone2015} However, simulations also show that cold fronts propagate to large radii at a roughly constant velocity with time \citep{Roediger2011,Roediger2012}, as gas at these radii becomes caught up in the pattern of motions driven by the cluster merger.

\begin{figure}
\begin{center}
\vbox{
\includegraphics[width=\linewidth]{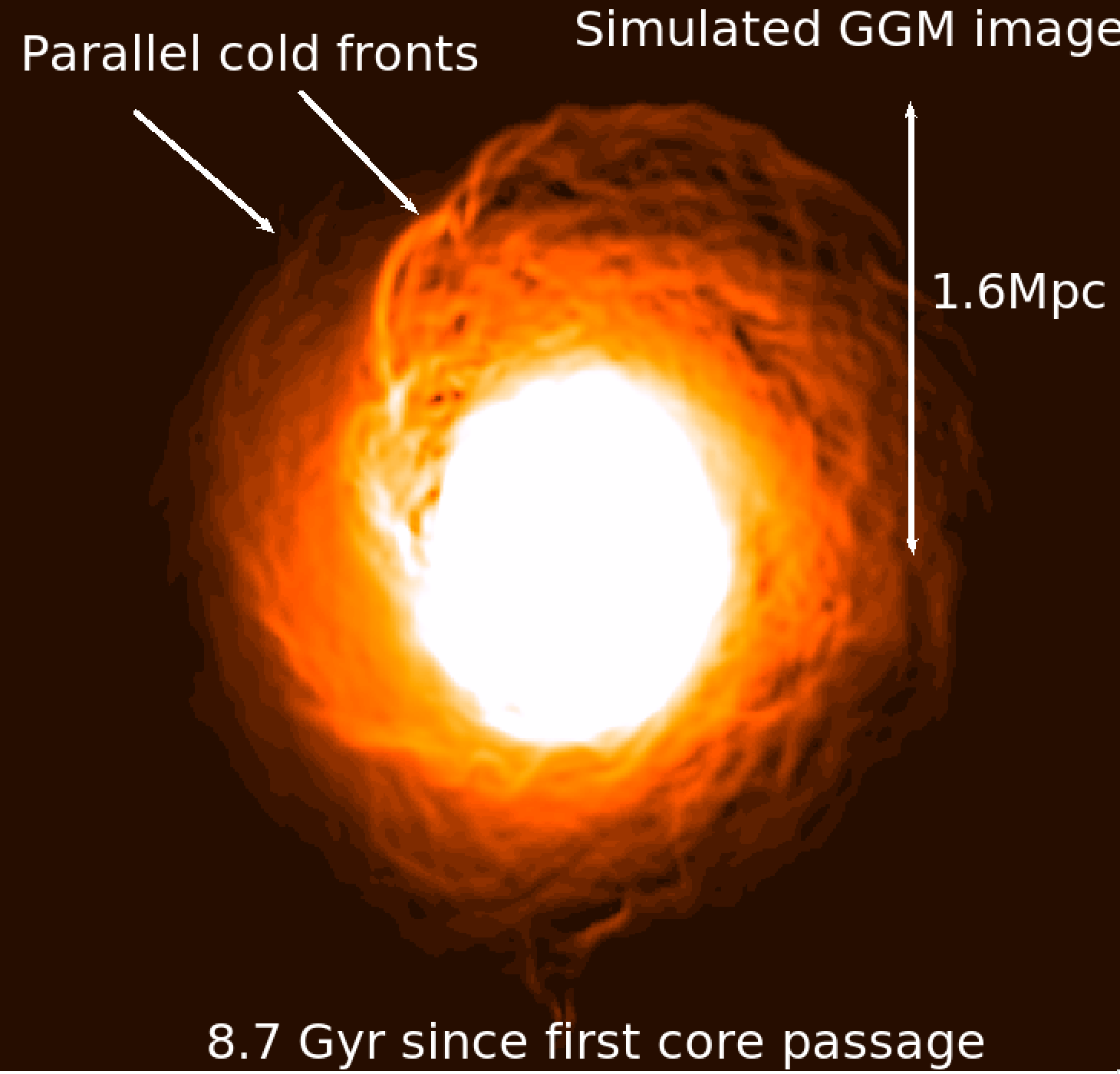}
\includegraphics[width=\linewidth]{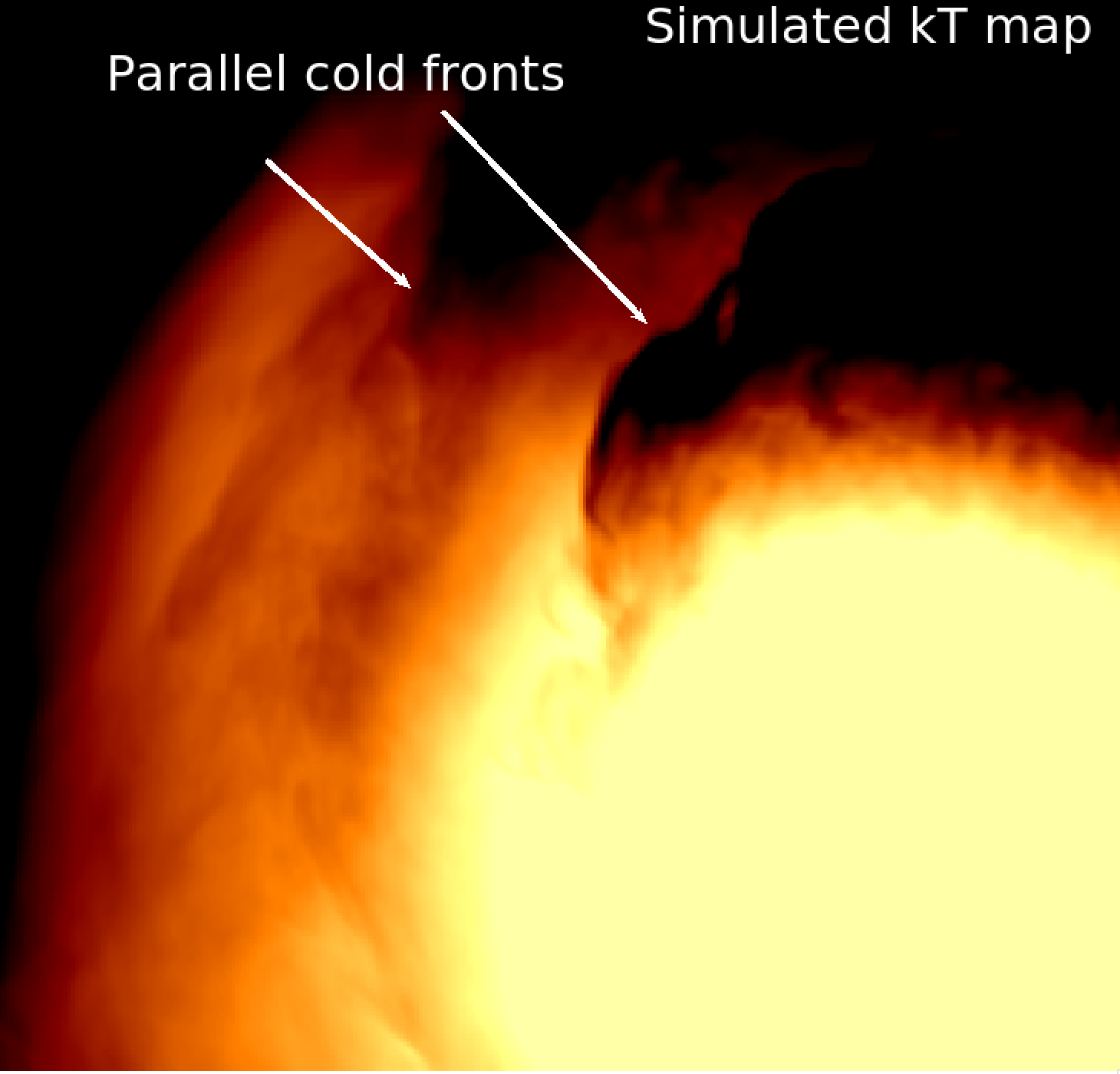}
}

\end{center}
\caption{\emph{Top}: Gaussian Gradient Magnitude filtered image of the sloshing simulation 8.7Gyr from the first core passage. We see that parallel cold fronts can form in the outskirts. In this simulation the outer cold front reaches 1.6Mpc. \emph{Bottom}: Zoom in of the temperature map for the parallel cold fronts in the cluster outskirts.
  }
\label{Simulations_plots}

\end{figure}

In Figure \ref{Simulations_plots} we show the results of a simulation of a galaxy cluster merger from \cite{Brzycki2019}, which includes dark matter, gas, and magnetic fields. The mass ratio of this merger is $R = 3$, and the larger cluster has a mass of $M_{200} = 6 \times 10^{14}~M_\odot$ (similar to the mass of the Perseus cluster found in \citealt{Simionescu2011} of $6.6^{+0.43}_{-0.46} \times 10^{14}~M_\odot$). Each cluster possesses a cool core and the initial ratio of thermal to magnetic pressure is $\beta = 200$. They approach each other with an initial impact parameter of $b = 1$~Mpc, which initiates large-scale gas motions which produce the cold fronts. The figure shows the state of the cluster at $t = 8.7$~Gyr past the first core passage of the subcluster. The top panel shows the GGM-filtered image of the projected X-ray emission from the cluster, revealing two cold fronts in parallel at a very large radius, up to $\sim$1.6~Mpc. These two cold fronts were produced shortly after core passage, near the core, and have expanded outward since, stabilized by the magnetic field despite the presence of turbulence and shock fronts. By comparison to simulations from previous works, it appears that the small mass ratio of the merger / relative large mass of the subcluster is essential to produce large cold fronts, since larger mergers produce faster gas motions which can more easily lift cold fronts to larger radii. However, the large impact parameter is also essential, since a lower-impact parameter merger would destroy the core. This may indicate that the Perseus cluster underwent one of these events in the distant past. Simulations more directly tailored to the Perseus cluster will be the subject of future works. 

\section{Conclusions}

We have analysed new XMM-Newton observations of the western outskirts of the Perseus cluster, following up evidence from ROSAT that there is an excess in X-ray surface brightness in this direction that appears to be due to a continuation of the gas sloshing seen in the core and around 700kpc from the core to the east. Our new XMM-Newton observations have found there to be a sharp edge, consistent in width with the mean free path, lying near the virial radius (1.7Mpc) to the west. Our analysis of the gas temperature using Suzaku spectra of the edge shows that there is also a sharp increase in the gas temperature from the more dense side to the lower density side. All of these properties make this edge consistent with being the largest cold front ever observed in a galaxy cluster. 

We also find another edge to the west, at a distance of 1.2Mpc from the core, which appears to run roughly parallel to the 1.7Mpc radius outer edge. The 1.2Mpc edge also has a width comparable to that of the mean free path, and has a temperature increase from the more dense side to the less dense side, making it consistent with being a cold front. 

We also use a numerical simulation of a binary galaxy cluster merger to show that such cold fronts can indeed be produced by sloshing motions in the core and expand outward to large radii, stabilized against turbulence by magnetic fields. The formation of these large fronts appears to require large-impact parameter, low-mass ratio mergers, which can drive fast gas motions without destroying the core completely. 

These results indicate that gas sloshing can continue out to extremely large radii in galaxy cluster, out to the virial radius. The sharpness of the edges shows that the cold fronts remain intact and do not disintegrate, even as they enter into the cluster outskirts where turbulence from infalling matter increases.

\section*{Acknowledgements}
SAW and MSM acknowledge support from the NASA XMM-Newton grant 80NSSC19K1698. Based on observations obtained with XMM-Newton, an ESA science mission with instruments and contributions directly funded by ESA Member States and NASA.%
\bibliographystyle{aasjournal}
\bibliography{PerseusLSS}

\newpage

\appendix
\section[]{Observations used}
\label{appendix_obs}

\begin{table*}
\begin{center}
\caption{XMM data used in this paper ordered by date taken. The four observations taken in 2019 extend the coverage of the mosaic to the west and are the focus of this paper.}
\label{obsdata}
\leavevmode
\begin{tabular}{ l|l | l | l | l| l} \hline \hline
 Obs ID & RA & Dec & Start Date & Exposure (ks) & Distance from core (arcmin)\\ \hline
                0820720401 &	       03 13 29.22 &	        +41 53 08.7 &	2019-03-04   &	               28.8 &	74.01\\
                0820720301 &	       03 13 16.44 &	        +41 35 51.8 &	2019-02-26   &	               25.0 &	73.28\\
                0820720201 &	       03 15 26.68 &	        +41 45 05.4 &	2019-02-24   &	               25.3 &	50.74\\
                0820720101 &	       03 15 16.88 &	        +41 24 17.2 &	2019-02-20   &	               25.4 &	51.06\\
                0673020401 &	       03 21 14.16 &	        +41 10 52.3 &	2012-03-01   &	               31.9 &	25.75\\
                0673020201 &	       03 23 07.63 &	        +41 12 35.6 &	2011-09-10   &	               39.0 &	41.78\\
                0673020301 &	       03 22 10.00 &	        +41 23 00.0 &	2011-08-19   &	               35.9 &	27.87\\
                0672770101 &	       03 28 00.10 &	        +41 29 56.5 &	2011-08-04   &	               16.9 &	92.28\\
                0554500801 &	       03 25 25.20 &	        +40 46 23.6 &	2008-08-19   &	               34.1 &	77.59\\
                0405410101 &	       03 21 04.26 &	        +41 56 05.0 &	2006-08-03   &	               30.9 &	28.72\\
                0405410201 &	       03 18 39.93 &	        +41 06 31.4 &	2006-08-03   &	               34.0 &	27.35\\
                0305690301 &	       03 19 50.59 &	        +41 53 34.3 &	2006-02-11   &	               27.2 &	22.8\\
                0305690401 &	       03 21 53.70 &	        +41 49 30.1 &	2006-02-11   &	               27.9 &	30.14\\
                0305690101 &	       03 18 02.69 &	        +41 16 60.0 &	2006-02-10   &	               27.9 &	23.96\\
                0305780101 &	       03 19 48.00 &	        +41 30 40.7 &	2006-01-29   &	              125 &	0.18\\
                0306680301 &	       03 13 01.97 &	        +41 20 01.2 &	2005-09-04   &	               63.4 &	76.72\\
                0305720101 &	       03 17 57.99 &	        +41 45 57.0 &	2005-09-01   &	               21.8 &	25.42\\
                0305720301 &	       03 22 15.99 &	        +41 11 28.0 &	2005-08-03   &	               28.3 &	33.95\\
                0204720201 &	       03 23 23.60 &	        +41 31 41.0 &	2004-02-04   &	               24.9 &	40.52\\
                0204720101 &	       03 21 38.59 &	        +41 31 43.0 &	2004-02-04   &	               17.9 &	20.87\\
                0151560101 &	       03 16 42.99 &	        +41 19 29.0 &	2003-02-26   &	               29.4 &	36.33\\
                0002942401 &	       03 15 01.40 &	        +42 02 09.0 &	2002-01-28   &	                7.9 &	61.83\\
                0085590201 &	       03 19 49.69 &	        +41 05 47.0 &	2001-02-10   &	               43.3 &	25\\
                0085110101 &	       03 19 48.16 &	        +41 30 42.1 &	2001-01-30   &	               60.8 &	0.2\\ \hline

\end{tabular}
\end{center}
\end{table*}

\begin{table*}
\begin{center}
\caption{Suzaku data used in this paper.}
\label{obsdata_Suzaku}
\leavevmode
\begin{tabular}{ l|l | l} \hline \hline
 Obs ID & RA & Dec \\ \hline
805117010  &	47.1647 &	41.6449 \\
805109010 &	47.3511 &	41.6379 \\
805116010 &	47.5391 &	41.63 \\
805108010 &	47.7251 &	41.6255 \\
805115010 &	47.9127 &	41.6204 \\
805107010 &	48.1001 &	41.6131 \\
805114010 &	48.2846 &	41.6054 \\
805106010 &	48.4696	 & 41.6016 \\
805105010 &	48.8462	 & 41.5856 \\
805104010 &	49.2188 &	41.5721 \\
805103010 &	49.5928  &	41.5523 \\ \hline
\end{tabular}
\end{center}
\end{table*}

\end{document}